\documentclass[11pt]{article}

 \newcommand{\be}{\begin{equation}}
 \newcommand{\ee}{\end{equation}}
 \newcommand{\ba}{\begin{eqnarray}}
 \newcommand{\ea}{\end{eqnarray}}
 \newcommand{\bl}{\begin{equation}\begin{array}{ll}}
 \newcommand{\el}{\end{array}\end{equation}}
 \newcommand{\bll}{\begin{equation}\begin{array}{lll}}
 \newcommand{\bdm}{\begin{displaymath}}
 \newcommand{\edm}{\end{displaymath}}
 \def\p{\partial}
 \def\f{\varphi}
 \def\ve{\varepsilon}
 \def\ra{\rightarrow}
\def\lim{\rightarrow}
\def\half{\frac{1}{2}}
\def\quart{\frac{1}{4}}

\def\be{\begin{equation}}
\def\ee{\end{equation}}
\def\bea{\begin{eqnarray}}
\def\eea{\end{eqnarray}}

\def\dif{\partial}

\begin{document}
\raggedbottom

\title{Integrable Low Dimensional Models for Black Holes \\
       and Cosmologies from High Dimensional Theories}

\author{ V.~de~Alfaro\thanks{vda@to.infn.it}~  and
A.T.~Filippov\thanks{filippov@thsun1.jinr.ru}~ \\
{\small \it {* DFTT and INFN, University of Turin,  Turin, I-10125}}\\
{\small \it {* Accademia delle Scienze di Torino, Turin, I-10131}}\\
{\small \it {$^+$ Joint Institute for Nuclear Research, Dubna, Moscow Region
RU-141980} }}

\maketitle

\begin{abstract}

We describe a class of integrable models of 1+1 and
1-dimensional dilaton gravity coupled to scalar fields.
The models can be derived from high dimensional supergravity
theories by dimensional reductions. The equations of motion
of these models reduce to systems of the Liouville equations
endowed with energy and momentum constraints. We construct
the general solution of the 1+1 dimensional problem in terms
of chiral moduli fields and establish its simple reduction
to static black holes (dimension 0+1), and cosmological models
(dimension 1+0). We also discuss some general problems of
dimensional reduction and relations between static and cosmological
solutions.
\end{abstract}

\section{Introduction}

It is now long time since 1+1 and 0+1 (or, 1+0) dimensional dilaton
gravity (DG) coupled to scalar matter fields proved to be a reliable
model for high dimensional (HD) black holes (BHs), cosmological models
(CMs), and branes (BRs).

The connection between high and low dimensions has
been demonstrated in different contexts of gravity and string
theory - symmetry reduction, compactification, holographic
principle, AdS/CFT correspondence, duality (see e.g. \cite{CGHS} -
\cite{Iv}). For spherically symmetric
configurations the description of static BHs, BRs and
CMs even simplifies to 1-dimensional DG -- matter models, often
analytically integrable (see e.g. \cite{A1}, \cite{Kiem} - \cite{A3}).

However, generally DG models coupled to matter are not integrable.
Recently A.T. Filippov has proposed a general class of integrable
1+1 dimensional DG models that
reduce to $N$ Liouville equations (a brief summary without proof is
published in Refs.\cite{A2}, \cite{A3}) and has found the general
analytic solution of the constraints. These models are
closely related to physically interesting solutions of higher
dimensional supergravity (SG) theories describing the low energy limit of
superstring theories.

A characteristic feature of the
static solutions of the models derived from string theory
is the existence of horizons with nontrivial scalar field distributions.
In fact, it is well known that in the Einstein - Maxwell theories
minimally coupled to scalar fields the spherical static horizons disappear
if the scalar fields have a nontrivial space distribution (this is the
`no-hair theorem'; we call `no horizon theorem' its local version).
However, the `no horizon' theorem is not true
for Einstein - Yang-Mills theories \cite{M1}
as well as for solutions of HD
            supergravity theories, see e.g. \cite{Kiem} -\cite{Iv}.

In Section~2 we demonstrate that BHs, BRs
and CMs derived from HD SG theories
may be described in terms of 1+1 and 1-dimensional DG theories.
In Sections~3,~4 we  discuss a new class of integrable DG theories
coupled to any number of scalar fields,
construct the general solutions and study their
main properties.
In Section~5 we briefly outline possible applications of the integrable
models and some unsolved problems. In the appendices we present
a selection of known results as well as
new ideas concerning dimensional reduction. We demonstrate that
a too naive reduction may result in loosing interesting
physical solutions. In particular, this gives a hint of
explanation for the fact that dimensional reduction to black holes
and cosmological models utilize very different approaches,
as one may conclude by inspecting relevant textbooks and reviews.

We also give a detailed derivation of the effective potential for
DG nonlinearly coupled to Abelian gauge fields.
The models with nonlinear coupling of gauge fields to DG were not
considered in the literature in full generality and the construction of
the effective potential for them is, to the best of our knowledge,
a new result. This result was mentioned in \cite{A2} and the proof was
given in \cite{VDA3}. Here we present a more transparent and more
general proof.

Note that in the lectures \cite{VDA3} the reader may find
a substantial part of the material presented here from
a different perspective. In fact, after completing
\cite{VDA3} we have found new results concerning relations
between BHs and cosmologies that significantly changed our
understanding dimensional reduction and the role of
the low dimensional integrable models in theories of
gravity and cosmology. Having this in mind, we decided
to completely rewrite \cite{VDA3}, preserving its formal
structure but deleting introductory or outdated material
while adding new original results (especially in
Sections~3,~4). We also attempted to be as clear and rigorous
as possible in formulating and interpreting our results
as well as in discussing their relations to other approaches.
For all these reasons, the present paper may be and should be
read independently of the lectures \cite{VDA3}.

\section{From HD to (1+1) dimensional dilaton gravity}

The HD theories which, under dimensional reductions, produce special
examples of integrable theories  come from the low energy limit of
the  theories described by 10-dimensional SGs. The bosonic part of the
SG of type II (corresponding to the type II superstrings) is
 \be
 {\cal L}^{(10)} =
{\cal L}_{NS-NS}^{(10)} \, + \, {\cal L}_{RR}^{(10)} \, . \label{eq:1}
\ee
Here it is sufficient to consider the first Lagrangian
(the second one gives similar 1+1 dimensional theories).  We have
\be
{\cal L}_{NS-NS}^{(10)} \, = \sqrt{-g^{(10)}} e^{-2\phi_s}
\biggl[ R^{(10)} +
4(\nabla \phi_s)^2 - {1\over 12} H_3^2\biggr]\, , \label{eq:2}
\ee
where $\phi_s$ is the dilaton, related to the string coupling
constant; $H_3 = dB_2$ is a 3-form; $g^{(10)}_{\mu \nu}$ and $R^{(10)}$
are the 10-dimensional metric and scalar curvature.

Among the many ways to reduce HD theories to low dimensions (LD)
we only mention those that may lead to integrable
theories. First, one may compactify a $D$-dimensional theory on
a $p$-dimensional torus $T^p$
and use the Kaluza - Klein - Mandel - Fock   (KKMF)
            mechanism\footnote{It is usually called the KK mechanism}.
            This introduces $p$ Abelian gauge fields and at least $p$ scalar fields.
Antisymmetric tensor fields ($n$-forms), which may be present in the HD theory,
will produce lower-rank forms and, eventually, other scalar fields. Thus we get
a theory of gravity coupled to matter fields (scalars, Abelian gauge fields
and, possibly, higher-rank forms) in a space of dimension $d=D-p$.
In the next step one reduces further its dimension by using some symmetry
of the $d$-dimensional
theory, most typically the spherical symmetry (the axial symmetry
leads to much more complex low dimensional theories and will not be considered
here). One thus arrives at a 1+1 DG theory coupled to scalar
and gauge fields.

The 1+1 dimensional theories so derived may describe
spherically symmetric evolution of the BHs (collapse of
matter) and of the universe (expansion of the universe).
Usually  the final step in the chain of dimensional
reductions in CMs is somewhat different from that in BH
physics since the CMs are normally obtained by reducing
the $d$-dimensional theory directly to dimension 1+0
(see e.g. \cite{Lu}, \cite{Lid}). Indeed, isotropy
and homogeneity of the universe require that the whole space should
have constant curvature $k=0,\,\pm 1$ and all these cosmologies
may be described by 1+0 dimensional DG theories
On the other hand, the spherical symmetry reduces HD gravity to
1+1 dimensional DG theories. In the case of the standard BHs coupled only
to Abelian gauge fields these 1+1 DG theories automatically reduce
to 1-dimensional DG.
In this way only a very special cosmology may emerge. It is
similar to the interior of the Schwarzschild black hole.

When there is scalar matter then, in general, there is no automatic reduction
from dimension 1+1 to 1+0 or 0+1 and attempts to use a naive reduction
by taking all geometric and matter fields depending on one variable only
($t$ or $r$, resp.) will result in loosing important solutions.
Moreover, in order to obtain the standard cosmological solutions
from the 1+1 dimensional
DG one has to modify the procedure of the dimensional reduction.
The modifications
that we discuss in a separate publication \cite{VDA2} give, as a bonus, some
unusual cosmological and static solutions. We will not discuss these matters
in detail, considering mainly naive reductions of the 1+1 DG.

 Returning to the dimensional reduction of SG theories, we only note that
one may use different sorts of dimensional reduction (KKMF, compactification
 on tori, etc.) but after several steps the resulting Lagrangian in dimension
 $d$ will typically depend on the curvature term $R^{(d)}$, a dilaton $\phi_d$, other
 scalar fields and Abelian forms (we do not consider reductions producing
non-Abelian forms as they do not give integrable theories even in low dimensions).
Thus, for our purposes the following expression for the effective Lagrangian
${\cal L}^{(d)}$ is sufficient:
\be
{\cal L}^{(d)} = \sqrt{-g^{(d)}} e^{-2\phi_d} \biggl( R^{(d)} + 4(\nabla
\phi_d)^2 - (\nabla \psi)^2
- X_0 - X_1 (\nabla \sigma)^2 - X_2 F_2^2 \biggr). \label{eq:3}
\ee
Here $F_2$ is a 2-form (an Abelian gauge field); the potentials
$X_a$ are functions of
$\phi_d$ and $\psi$. Actually, the Lagrangian should depend on
several $F_2$-fields, several $\psi$-fields, and may depend on several
$\sigma$-fields  as well as on higher - rank forms\footnote{
Typically, in dimensional reductions of supergravity
by toroidal compactifications and the KKMF mechanism
there is no $X_0$ terms but higher-rank forms usually
appear. Nevertheless, if the compactified $(D-d)$-dimensional
manifold has nonzero Ricci curvature such a term will appear
as dictated by (\ref{a2}) (see Appendix 6.1)}.
However, after further reduction to dimension 1+1  only 2-forms and scalar
fields will survive (the 2-forms can also be excluded by writing
an effective potential depending on electric or magnetic charges).

The $d$-dimensional theory can be further reduced to dimension 1+1 by spherical
symmetry. Before and after doing so one may transform this Lagrangian by
the Weyl conformal transformation, $g_{\mu\nu} \Rightarrow \tilde{g}_{\mu\nu}
\equiv \Omega^2 g_{\mu\nu}$,
where $\Omega$ depends on the dilaton. Expressing $R$ in terms of the
new metric,
\be
R^{(d)} = \Omega^2 \biggl[ \tilde{R}^{(d)} + 2(d-1) {\tilde{\nabla}}^2
\ln{\Omega} -
(d-1)(d-2) (\tilde{\nabla} \ln{\Omega})^2 \biggr] \, , \label{eq:3a}
\ee
one can easily find the new expression for the Lagrangian. For $d > 2$
the multiplier $e^{-2\phi_d}$ can be eliminated by choosing an appropriate
function $\Omega(\phi_d)$ and thus the Lagrangian can be written in the
`Einstein frame' (as distinct from the `string frame' expressions above).
If $d=2$ the dilaton multiplier cannot be removed but, instead, one can
remove the dilaton gradient term.

For the spherically symmetric solutions of the theory (\ref{eq:3})
it is more convenient to remove the dilaton factor by a Weyl transformation
and rewrite the action (\ref{eq:3}) in the Einstein frame,
\be
{\cal L}^{(d)}_E = \sqrt{-g^{(d)}} \biggl[ R^{(d)} - (\nabla \chi)^2
- (\nabla \psi)^2
- X_0 e^{a_0 \chi} - X_1 e^{a_1 \chi } (\nabla\sigma)^2 -
 X_2 e^{a_2 \chi } F_2^2 \biggr], \label{eq:3b}
\ee
where $\chi \propto \phi_d$ and $a_k$ are known constants depending on $d$.
 Then we parameterize the spherically symmetric metric\footnote{
 Actually, we here call spherical symmetry a somewhat more general
 symmetry. Below, $d\Omega_{(d-2)}^2 (k)$ is the metric on the surface
 of the generalized $(d-2)$-dimensional `unit sphere' $S^{(d-2)}(k)$ that
is the
 flat space for $k=0$, pseudo-sphere for negative curvature $k=-1$ and the
 standard sphere for positive curvature $k=1$.
 The cylindrical symmetry also reduces the $d$-dimensional theory (\ref{eq:3b})
 to a 1+1 dimensional dilaton gravity but the equations of motion are more
 cumbersome than in the spherical case.}
by the general 1+1 dimensional metric $g_{ij}(x^0,x^1)$ and the new dilaton
$\f(x^0,x^1)$ ($\nu \equiv 1/n$, $n \equiv d-2$),
\be
ds^2=g_{ij}\, dx^i\, dx^j \,+\, {\f}^{2\nu} \,
d\Omega_{(d-2)}^2 (k) \, , \label{eq:4}
\ee
introduce appropriate spherical symmetry conditions for the
fields, which from now on will be functions of the variables $x^0$ and $x^1$
($t$ and $r$), and integrate out the other (angular)
variables from the action.

In concrete computations, it is often more convenient to use the diagonal
spherically symmetric metric
\be
ds^2 = e^{2\alpha} dr^2 + e^{2\beta} d\Omega^2_{(d-2)}(k)
- e^{2\gamma(t)} dt^2 \, ,
\label{eq:4a}
\ee
 where $\alpha , \beta , \gamma$ are functions of $r$ and $t$.
It should be emphasized that one may pass to this metric only after deriving
the equation of motion with the general metric (\ref{eq:4}).
Otherwise one of the equations will be lost.
 Note that the 2-dimensional dilaton
 is essentially the metric coefficient in the HD theory:
            $\f(r,t) = e^{\beta(r,t) /\nu}$.

Applying, in addition, the Weyl transformation that removes the dilaton
gradient term we may obtain the effective 1+1 dimensional action
\ba
{\cal L}^{(2)} = \sqrt{-g} \biggl[ \f R + k\,n(n-1) \f^{-\nu} -
X_0 e^{a_0 \chi } \f^{\nu} -
X_2 e^{a_2 \chi} \f^{2-\nu} F^2 -
\nonumber \\
 - \f \biggl( (\nabla \chi )^2
+ (\nabla \psi)^2 + 2 X_1 e^{a_1 \chi } (\nabla \sigma)^2
\,\biggr) \, \biggr] . \label{eq:5}
\ea
Here $R\equiv R^{(2)}$, $\f$ is the 2-dilaton field;
the scalar fields $\psi$ may have different origins -- former
dilaton fields, KKMF scalar fields, reduced $p$-forms, etc. The potentials
$X_m$ depend on the scalar fields $\chi$ and
$\psi$, which from now on will be called scalar matter fields. Also the
field $\sigma$ may be regarded as a matter field but it plays a special
role that will be discussed later. Notice that the potentials
        $X_m$ are positive.

In general, these theories are not integrable,
meaning that it is impossible to find the analytic
form of the general solution of the equations of
motion. The reason is that EMs are highly
nonlinear. The well known exception is the theory
in which $k=0$, $X_0 = 0$, $F^2 =0$ and
$\sigma = 0$. Although formally the EMs are still
nonlinear, using the light cone metric one can show that
the scalar fields satisfy the linear Euler -- Darboux
equation, for the general solution of which one can
write a rather complex integral representation
\cite{Szekeres}. These solutions describe plane waves of scalar
matter coupled to gravity. However, the expression
for the metric is very difficult to analyze.

As mentioned above, the CMs are usually obtained by a different
dimensional reduction, where the metric is written in the form
\be
ds^2 = -e^{2\gamma(t)} dt^2 + e^{2\beta(t)} d\Omega^2_{(d-1)}(k)\, .
 \label{eq:5a}
\ee
Also, in this case somewhat different reductions of the fields are
of interest because terms generated by the higher rank forms are important.
However, after reduction  the higher rank forms give rise to scalars
of the $\psi$ or $\sigma$ type.

More important is the fact that the 1-dimensional static and cosmological
 solutions are differently embedded into the 1+1 DG theories. For this reason,
 a relation between the standard cosmological models and black holes,
and, more generally, between different 1-dimensional reductions of the
 1+1 DG is by no means obvious.
 One problem is of course to check that the solutions of the dimensionally
 reduced theories satisfy the original HD equations. The second problem is
 to classify and enumerate all possible reductions of both types.
 To clarify this matter one has to carefully
 review the whole procedure of dimensional reduction because:
1.~total derivative terms which are usually omitted without discussion
may give non trivial contributions
to low dimensional Lagrangians and 2.~gauge fixing looking quite innocent
if you work in one and the same dimension may lead to a loss of
information when you perform a
dimensional reduction. The effects of a careless approach to these two problems
may be twofold: 1.~ some LD solutions of the HD equations may be lost,
and 2.~ some solutions of the LD equations (obtained by too naive reductions)
may not satisfy the HD equations (these effects are
 demonstrated by simple examples in \cite{VDA2}).

\section{Integrable 2-dimensional DG coupled to matter}

The Lagrangian of the general 1+1 DG coupled to Abelian gauge fields
$F^{(a)}_{ij}$ and to scalar fields $\psi_n$ is
\bea
{\cal L}^{(2)} = \sqrt{-g} \, [ \, U(\f) R(g) +
V(\f) + W(\f)(\nabla \f)^2 + \nonumber \\
+ X (\f,\psi, F_{(1)}^2, ..., F_{(A)}^2) + Y(\f,\psi)+ \sum_n Z_n(\f,\psi)
(\nabla \psi_n)^2 \, ] \, . \label{eq:6}
\eea
Here $g_{ij}$ is a generic 1+1  metric with signature (-1,1) and
$R$ is the Ricci curvature; $U(\f)$, $V(\f)$, $W(\f)$ are
arbitrary functions of the dilaton field; $X$,
$Y$ and $Z_n$ are arbitrary functions of the
dilaton field and of $N-2$ scalar fields $\psi_n$ ($Z_n < 0)$;
$X$ also depends on $A$ Abelian gauge fields
 $F_{(a)ij} \equiv F^{(a)}_{ij}$,
 $F_{(a)}^2 \equiv g^{ii'} g^{jj'} F^{(a)}_{ij} F^{(a)}_{i'j'}$.
Notice that in dimensionally reduced theories
both the scalar fields and the Abelian gauge fields are
non-minimally coupled to the dilaton (i.e. the corresponding potentials
depend on the dilaton).

The equations of motion (EM) of the theory (\ref{eq:6})
can be solved for arbitrary potentials $U$, $V$, $W$ and $X$ if
 $Y \equiv 0$, $\dif_{\psi}X \equiv 0$, and $\nabla \psi_n \equiv 0$.
 Then the theory is equivalent to the pure
 dilaton gravity (the one for which the potential $X$, $Y$, $Z$ vanish)
 that is known to be an integrable topological theory
(for the simplest explicit solution in case of $X$ linear in $F^2$
see e.g \cite{A1} and references therein as well as the recent review
\cite{Kummer}, where on may find further references).

The properties of the general DG (\ref{eq:6}) are much more complex.
To simplify further consideration we solve the equations for $F^{(a)}_{ij}$
and construct the    effective action (see Appendix 6.3)
\be
{\cal L}_{\rm eff}^{(2)} =\sqrt{-g}\left[ \f R(g) + V_{\rm eff}(\f,\psi) +
\sum_n Z_n(\f,\psi)\, g^{ij} {\dif}_i \psi_n {\dif}_j \psi_n \right]\,.
\label{eq:7}
\ee
Here the effective potential $V_{\rm eff}$ (below we omit the subscript)
depends also on charges introduced by solving the equations for the Abelian
fields. We have used in (\ref{eq:7}) a Weyl transformation to exclude the
kinetic term for the dilaton and also choose the simplest, linear parameterization
for $U(\f)$\footnote{ If $U^{\prime}(\f)$ has zeroes, this parameterization,
as well as the more popular exponential one, $U = e^{\lambda \f}$, is valid
only locally, i.e. between two consecutive zeroes.}.

If $V_{\rm eff} = V(\f)$ and $Z_n \equiv 0$
the theory (\ref{eq:7}) is the pure dilaton gravity
that is integrable with arbitrary potential $V(\f)$.
If $Z_n \neq 0$ the theory may be integrable with
very special potentials $V_{\rm eff}(\f,\psi)$ and
$Z_n(\f, \psi)$. Roughly speaking, the two dimensional
models may be integrable if either the potentials $Z_n$
can be transformed to constants or the potential
$V_{\rm eff}$ is zero. In the first case $V_{\rm eff}$
should have very special form what is described below.
In the second case,
for the theory to be explicitly analytically
integrable the potentials $Z_n$ must be very
special functions\footnote{
The theories proposed in \cite{FI} are explicitly
integrable. The theories with $Z_n = -\f$ are essentially
equivalent to the one implicitly solved
in \cite{Szekeres}. String theory may generate a more
general scalar coupling in ${\cal L}_{\rm eff}^{(2)}$ related
to $\sigma$-models. Such theories
were considered in many papers in connection to studies
of the cosmological singularity (see \cite{Venezia}, where
a discussion of the main results and further references
may be found.)}.
The best studied examples of the integrable
models belonging to the first class (constant $Z_n$) are the
CGHS model ($V_{\rm eff} = g_0$ ) and the JT model
($V_{\rm eff} = g_1 \f$). A generalization
($V_{\rm eff} = g_{+}e^{g\f} + g_{-}e^{-g\f}$)
was proposed in \cite{A1}.

Here we will mainly discuss a much more general class of
integrable 1+1 DG models with minimal coupling to scalar fields
defined by the Lagrangian (\ref{eq:7})
with the following potentials:
\be
Z_n = -1; \;\;\;\;  |f|V_{\rm eff} = \sum_{n=1}^N 2g_n e^{q_n} \, . \, \label{eq:8}
\ee
Here $f$ is the light cone metric, $ds^2 = -4f(u, v) \, du \, dv$, and
\be
q_n \equiv F + a_n \f + \sum_{m=3}^{N} \psi_m a_{mn} \equiv
\sum_{m=1}^{N} \psi_m a_{mn} \, , \label{eq:9}
\ee
where $\psi_1 + \psi_2 \equiv \ln{|f|} \equiv F$
($f \equiv \varepsilon e^F$, $\varepsilon = \pm 1$),
$\psi_1 - \psi_2 \equiv \f$ and thus
$a_{1n} = 1 + a_n$, $a_{2n} = 1 - a_n$
By varying the Lagrangian (\ref{eq:7}) in $N-2$ scalar fields, dilaton and in
$g_{ij}$ and then passing to the light cone metric we find $N$
equations of motion for $N$ functions $\psi_n$,
\be
\epsilon_n \p_u \p_v \psi_n  =  \sum_{m=1}^{N} \varepsilon g_me^{q_m} a_{mn}
\, ; \,\,\,\,
\epsilon_1 = -1, \,\,\, \epsilon_n = +1, \,\, {\rm if} \,\, n \geq 2 \, ,
\label{eq:10}
\ee
as well as two constraints,
\be
C_i \equiv f {\dif}_i ({\dif}_i \f /f) + \sum_{n=3}^N ({\dif}_i \psi_n)^2 =
0 ,\,\,\,\,\,  i = (u, v) . \,\, \label{eq:11}
\ee

 For arbitrary coefficients $a_{mn}$ these EMs are not integrable.
 However, as proposed in \cite{A2} - \cite{A3}, the equations (\ref{eq:10}) are
 integrable and the constraints (\ref{eq:11}) can be solved if the $N$-component
 vectors $v_n \equiv (a_{mn})$ are pseudo - orthogonal.
 Then (\ref{eq:10}) reduce to $N$ independent, explicitly integrable
 Liouville equations for $q_n$,
\be
{\dif}_u {\dif}_v q_n - {\tilde{g}}_n e^{q_n} =0 , \label{eq:12}
\ee
where ${\tilde{g}}_n = \ve \lambda_n g_n$, $\lambda_n = \sum \epsilon_m
a_{mn}^2$, and $\ve \equiv f/|f|$ (note that the equations for $q_n$
depend on $\epsilon_n$ only implicitly, through the normalization factor
$\lambda_n$)\footnote{Here we suppose that $\lambda_n \neq 0$ and $g_n \neq 0$.
Otherwise the solution of the constraints should be modified in a fairly
 obvious way. We also denote $\gamma_n \equiv {\lambda_n}^{-1}$.}.

 The expression for the original fields in terms
 of the Liouville fields $q_n$
 may be found by using the orthogonality relations
 for $a_{mn}$ ($m \neq n$) and the definition of
 $\lambda_n \equiv \gamma_n^{-1}$ (for $m=n$)
 combined in the equation:
    \be
\sum_{k=1}^{N} \epsilon_k a_{km} a_{kn} = \lambda_n \delta_{mn} \equiv
\gamma_n^{-1} \delta_{mn} \ .
\label{eq:12b}
    \ee
More convenient is to use the following sum rules
that follow from (\ref{eq:12b}):
    \be
\sum \gamma_n =0 \ ; \,\,\, \sum \gamma_n a_n = -\half \ ; \,\,\,
\sum \gamma_n a_n^2 = 0 \ ; \,\,\, \sum a_{mn} \gamma_n = 0 \ , \,\,  m \geq 3 \ .
\label{eq:12c}
    \ee
Now, using (\ref{eq:12b}) -- (\ref{eq:12c}) we can invert the
definition (\ref{eq:9}) and get
\be
\psi_n = \epsilon_n \sum a_{nm} \gamma_m q_m \ , \,\,\,\,\,
F = -2 \sum \gamma_n a_n q_n \ , \,\,\,\,\,
\f = -2 \sum \gamma_n q_n \ .
\label{eq:12aa}
\ee

Also, using the orthogonality relations (\ref{eq:12b})
it can be proven that one and only one norm $\gamma_n$ is negative.
We thus choose $\gamma_1 < 0$ while other $\gamma_n$ are
positive. In physically motivated models the parameters
$a_{mn}$, $\gamma_n$ and $g_n$ may satisfy some further
relations. For example, the signs of $\gamma_n$ and
$g_n$ may be correlated so that $g_n / \gamma_n < 0$.
However, such relations do not follow from the orthogonality
conditions and we ignore them in our discussion\footnote{
In fact, the coupling constants have nothing to do with
integrability and may be arbitrarily chosen.}.

The most important fact is that the constraints can be explicitly solved.
First, we write the solutions of the Liouville equations
(\ref{eq:12}) in the form suggested by the conformal symmetry properties
of the Liouville equation \cite{Gervais},
\be
e^{-q_n /2} = a_n(u) b_n(v) - \half {\tilde{g}}_n \bar{a}_n(u) \bar{b}_n(v) \equiv X_n(u,v) \ ,
\label{eq:13}
\ee
where the chiral fields $a_n(u)$, $b_n(v)$, $\bar{a}_n(u)$
and $\bar{b}_n(v)$ satisfy the equations (do not mix $a_n(u)$ with $a_n$ used above)
\be
a_n(u) \bar{a}_n^{\prime}(u) - a_n^{\prime}(u) \bar{a}_n(u) = 1 , \,\,\,\,\,\,
b_n(v) \bar{b}_n^{\prime}(v) - b_n^{\prime}(v) \bar{b}_n(v) = 1 .
\label{eq:13a}
\ee
Using (\ref{eq:13a}) we can express $\bar{a}$ and $\bar{b}$
in terms of $a$ and $b$ and thus write $X_n$ as
\be
X_n(u,v)  = a_n(u) b_n(v) \biggl[ 1 - \half {\tilde{g}}_n
\int {du \over a_n^2(u)} \int {du \over b_n^2(v)} \biggr] \, .
\label{eq:13b}
\ee
It is not so straightforward but not very difficult to
rewrite the constraints (\ref{eq:11}) in the form\footnote{
If some $\tilde{g}_n$ vanish, the corresponding fields
$q_n$ should be moved to the right-hand side of
the equations (\ref{eq:11}). It is not difficult
to explicitly solve these generalized constraints.}
\be
  C_u \equiv 4\sum_{n=1}^N \gamma_n {{{a_n}^{\prime \prime} (u)} \over {a_n(u)}} = 0
\, , \,\,\,\,\,\,
C_v \equiv 4\sum_{n=1}^N \gamma_n {{{b_n}^{\prime \prime} (v)} \over {b_n(v)}} = 0 \, .
\label{eq:14}
\ee

Using the relation $\sum \gamma_n = 0$
we can explicitly solve these constraints.
First, it can be shown that the constraints are equivalent to
the equations
\be
{{a_n^{\prime} (u)} \over {a_n(u)}} = \mu_n (u) -
{{\sum \gamma_n {\mu_n}^{\prime}(u)} \over {2\sum \gamma_n \mu_n(u)}} \, ,
\,\,\,\,\,\,\,
{{b_n^{\prime} (v)} \over {b_n(v)}} = \nu_n (v) -
{{\sum \gamma_n {\nu_n}^{\prime}(v)} \over {2\sum \gamma_n \nu_n(v)}} \, ,
 \label{eq:15}
\ee
where $\mu_n(u)$ and $\nu_n(v)$ are arbitrary functions satisfying
the constraints
\be
\sum \gamma_n {\mu_n}^2(u) = 0 \, , \,\,\,\,\,\,\,
\sum \gamma_n {\nu_n}^2(v) = 0 \, .
\label{eq:15a}
\ee

Now, by integrating the first order differential equations
(\ref{eq:15}) for $a_n(u)$ and $b_n(v)$
we find the general solution of the $N$-Liouville DG in terms
of the chiral moduli fields $\mu_n(u)$ and $\nu_n(v)$
satisfying (\ref{eq:15a}):
\be
a_n(u) = {|\sum \gamma_m \mu_m|}^{-\half} \exp{\int du \, \mu_n(u)} , \,\,\,
b_n(v) = {|\sum \gamma_m \nu_m|}^{-\half} \exp{\int dv \, \nu_n(v)} .
\label{eq:15b}
\ee

The moduli fields $\mu_n(u)$ and $\nu_n(v)$ are not
independent due to the constraints (\ref{eq:15a}).
In addition, we may use coordinate transformation
$u \rightarrow U(u)$ and $v \rightarrow V(v)$
to choose two gauge conditions. We will show in a moment
that writing
    \be
\mid \sum \gamma_n \mu_n(u)\mid \equiv U^{\prime}(u) \ , \,\,\,\,\,
\mid \sum \gamma_n \nu_n(v)\mid \equiv U^{\prime}(v)
\label{eq:15c}
    \ee
is indeed equivalent to choosing (U,V) as a new coordinate
system. With this aim, we first write
\be
A_n \equiv \exp \int du \mu_n(u) \ , \,\,\,\,\,
B_n \equiv \exp \int dv \mu_n(v) \ .
\label{eq:15d}
\ee
Using (\ref{eq:15b}) and (\ref{eq:15c}) we may rewrite
eq.(\ref{eq:13a}) in terms of $A_n(U)$, $B_n(V)$:
\be
Y_n(U,V) = A_n(U) B_n(V) \biggl[ 1 -
\half {\tilde{g}}_n \int {dU \over A_n^2(U)} \int {dV \over B_n^2(V)} \biggr] \, ,
\label{eq:15e}
\ee
where we have defined
\be
Y_n(U,V) \equiv  [U^{\prime}(u) V^{\prime}(v)]^{\half} X_n(u,v)
\label{eq:15ea}
\ee

Now, with the above definitions
we may derive the metric $f$,
\be
f (u,v) = U^{\prime}(u) V^{\prime}(v) \prod_{n=1}^{N} [Y_n(U,V)]^{4\gamma_n a_n} \equiv
\bar{f}(U,V) U^{\prime}(u) V^{\prime}(v) ,
\label{eq:15f}
\ee
the dilaton $\f$, and the scalar fields $\psi_m$ ($m\geq 3$):
\be
e^{\f} = \prod_{n=1}^{N} [Y_n(U,V)]^{4\gamma_n} , \,\,\,\,\,
e^{\psi_m} = \prod_{n=1}^{N} [Y_n(U,V)]^{-2a_{mn} \gamma_n} .
\label{eq:15g}
\ee
We see that $ds^2 = -4f(u,v) \equiv -4 \bar{f}(U,V) dU dV$
and thus everything is expressed in terms of the new
coordinates $(U,V)$.

The constraints (\ref{eq:15a}) for the moduli parameters can
easily be solved. However, it may be more convenient
and instructive to introduce new moduli that are
unit $(N-1)$-vectors (recall that $\gamma_1 <0$ and
$\gamma_k > 0$ for $k \geq 2$):
\be
\hat{\xi}_k(u) \equiv {{\mu_k (u) \surd \gamma_k}\over {\mu_1 (u) \surd |\gamma_1}|} \ ,
\,\,\,\,\,\,\,\,\,
\hat{\eta}_k(v) \equiv {{\nu_k (v) \surd \gamma_k}\over {\nu_1 (v) \surd |\gamma_1}|} \ ,
\,\,\,\,\,\,\,\,\,  k=2,...,N \ .
\label{eq:15h}
\ee
These vectors moving on the surface of the unit sphere
$S^{(N-2)}$ determine the solution up to a choice of the
coordinate system, which can be fixed by the above gauge
conditions (\ref{eq:15c}). They now look as follows:
\be
U^{\prime}(u) = |\gamma_1 \mu_1 (u)| (1 - \cos \theta_{\xi}(u))  , \,\,\,\,\,\,\,
V^{\prime}(v) = |\gamma_1 \nu_1 (v)| (1 - \cos \theta_{\eta}(v)) ,
\label{eq:15k}
\ee
\be
\cos \theta_{\xi}(u)  = \sum_{k=2}^N \hat{\gamma}_k \hat{\xi}_k(u) , \,\,\,\,\,\,\,
\cos \theta_{\eta}(v) = \sum_{k=2}^N \hat{\gamma}_k  \hat{\eta}_k(v) ,
\label{eq:15l}
\ee
where $\hat{\gamma}$ is the constant unit $(N-1)$-vector,
$\hat{\gamma}_k = (\gamma_k / |\gamma_1|)^{\half}$.

We thus have the general solution of the 1+1 dimensional dilaton gravity
coupled to any number of scalar fields.
It is explicitly expressed in terms
of a sufficient number of arbitrary chiral fields
and thus may be regarded as a highly nontrivial
generalization of the D'Alembert solution for massless scalar fields.
Using this solution one
may solve the Cauchy problem and study the evolution of cosmological or
black hole type solutions, etc. The representation of the general solution
in terms of the chiral fields $a_n(u)$ and $b_n(v)$ may give us a good starting
point in attempts to quantize our $N$-Liouville DG.
Even more useful may be the chiral moduli fields $\mu_n(u)$ and $\nu_n(v)$
(or ${\hat{\xi}}_k(u)$ and ${\hat{\eta}}_k(v)$).
In terms of these moduli fields the dimensional reduction of the
solutions becomes very transparent and this may simplify the derivation and
 physical interpretation of the evolution of one dimensional solutions
 and suggest new approaches to quantization  based on analogy with
the simple  1-dimensional case.

\section{Integrable 1-dimensional DG coupled to matter}

The naive reduction from dimension 1+1 to 0+1 (1+0) in the light cone
coordinates $(u, v)$ is very simple.
 Supposing that $\f=\f(\tau)$, $\psi_n =\psi_n(\tau)$, where
 $\tau=a(u)+b(v)$, we find from the 1+1 dimensional EMs
 (see Appendix)
 \be
 f(u,v) = \mp h(\tau) \, a'(u) \, b'(v), \quad
 ds^2 = -4f(u,v) \, du \, dv = \pm 4h(\tau) da \,db.
 \label{eq:16}
 \ee

Defining the space and time coordinates $r = a \pm b$ and
$t = a \mp b$ we find
\be
    ds^2 = h(\tau) (dr^2 - dt^2) , \quad {\rm where} \quad
\tau = r \,\, {\rm or} \,\,  \tau = t .
\label{eq:17}
\ee
Thus the reduced solution may be static or cosmological\footnote{Of course,
in 2-dimensional theories this distinction is not important.
However, when we know the higher dimensional theory from which our DG
originated, we can reconstruct the higher dimensional metric and thus
find the higher dimensional interpretation of our solutions. }.

This is not the most general way for obtaining 0+1 or 1+0 dimensional
theories from higher dimensional ones. Not all possible reductions can be
derived by this naive
dimensional reduction of the 1+1 gravity.
For example, the cosmological solutions corresponding to the reductions
(\ref{eq:5a}) should be derived by a more complex dimensional reduction
of the 1+1 dimensional DG, which will be discussed in \cite{VDA2}.

The 0+1 EMs are described
by the Lagrangian ($h = \ve e^F$, $\ve = \pm 1$) \cite{A1}:
\be
{\cal L}^{(1)} =-{1\over l(\tau)} \biggl[ \dot \f \dot F +\sum_n
Z_n(\f, \psi) \dot \psi_n^2 \biggr] +l(\tau)\, \ve e^F \, V(\f , \psi) ,
\label{eq:18}
\ee
where $l(\tau)$ is the Lagrange multiplier (related to the general metric
$g_{ij}$).

The integrable  2-dimensional $N$-Liouville theories are also
integrable in dimension 0+1. Moreover, as we can solve the Cauchy problem
in dimension 1+1, we can study the evolution of the initial configurations
to stable static solutions,
e.g. BHs, which are special solutions of the 0+1 reduction.
However, the reduced theories can be explicitly solved for much more general
potentials $Z_n$ and $V$,
including those of realistic HD theories
of branes, black holes and cosmologies.
In this case, the 1-dimensional model (\ref{eq:18}) is
embedded in the nonintegrable 1+1 dimensional
field theory\footnote{
Unfortunately, we do not know any realistic HD theory
with non vanishing potentials
$Z_n = -\f$ that can be reduced to an integrable
theory in dimension 1+1.}.
Nevertheless, even in this case, integrable
1+1 dimensional theories may be used for approximate,
qualitative description of processes leading to
creation of BH as well as for studies of inhomogeneous
cosmologies etc.

Suppose that for $N-2$ scalar fields $\psi_n$ ($n = 3,...,N$) the
ratios of the $Z$-potentials are constant so that we can write
$Z_n = -\zeta_n /\phi^{\prime}(\f)$.
Let all the potentials $Z_n$ and $V$ be
independent of the scalar fields $\psi_{N+k}$ with $k = 1,...,K$.
Then we first remove
the factor $\phi^{\prime}(\f)$ by defining the new Lagrange multiplier
$\bar{l} = l(\tau) \phi^{\prime}(\f)$ and absorb the
factors $\zeta_n > 0$ in the corresponding scalar fields. So we
introduce the new dynamical variable $\phi$ instead of $\f$. Now we can
solve the equations for the $\sigma$-fields and construct the effective
Lagrangian:
\be
{\cal L}_{\rm eff}^{(1)} =-{1\over \bar{l} }\biggl[ \dot\phi \dot F - \sum_{n=3}^N
 \dot \psi_n^2 \biggr] + \bar{l} \biggl[ \ve e^F  V_{\rm eff}(\phi, \psi) +
 V_{\sigma} (\phi, \psi) \biggr] .
\label{eq:19}
\ee
Here $V_{\rm eff} = V/\phi^{\prime}(\f)$ and $V_{\sigma} =
        \sum_k C_k^2 {(Z_{N+k} \, \phi^{\prime}(\f))}^{-1}$
is the effective potential derived by
solving the EMs for the $\sigma$-fields
$\psi_{N+k}$ (of course, now $\f$ should
be expressed in terms of $\phi$).
If the original potentials in eq.~(\ref{eq:18}) are such that
$(Z_{N+k} \, \phi^{\prime}(\f))^{-1}$ and $V / \phi^{\prime}$
can be expressed in terms of sums of exponentials of linear combinations
of the fields $\phi$ and $\psi$, then there is a chance that the 1-dimensional
theory can be reduced to the $N$-Liouville theory
(or the Toda theory if $V_{\sigma} \neq 0$).

It should be emphasized that potentials $V(\f, \psi)$, which are exponential
(in $\f$) in the original 1+1 dimensional theory,
may become not exponential in the effective
1-dimensional theory (\ref{eq:19}), and vice versa. For example,
this is evident in the realistic theories where $Z_n = - \f$
and thus $\phi(\f) = \ln \f$. This example shows why integrable
1-dimensional theories may be naturally embedded in nonintegrable
2-dimensional theories. Of course, we may embed the theory (\ref{eq:19})
in the integrable 2-dimensional theory with the dilaton $\phi$, the potential
$V_{\rm eff}(\phi, \psi)$ and $Z_n = -1$ but the relation of such a theory
to the original HD (and 2-dimensional) theory is not quite clear.

Now let us forget about these subtleties and consider the 1-dimensional
Lagrangian (\ref{eq:18}) obtained by dimensional reduction from the
2-dimensional integrable $N$-Liouville theory (with $Z_n = -1$).
We thus have
\be
{\cal L}^{(1)} = {1\over l} \bigl( - \dot \psi_1^2 +
 \sum_{n=2}^N \dot \psi_n^2 \bigr) +l \sum_{n=1}^N 2\ve g_n e^{q_n} =
 \sum_{n=1}^N \bigl({1\over l} {\gamma_n} \dot q_n^2 + 2l \ve g_n e^{q_n} \bigr)
 \, ,
\label{eq:20}
\ee
 where $q_n$ is defined by eq.~(\ref{eq:9}) and  $a_{mn}$
 satisfy pseudo orthogonality conditions
 (\ref{eq:12b}). Then EMs are
reduced to $N$ independent Liouville equations whose solutions
 have to satisfy the energy constraint that can be derived
 by varying ${\cal L}$ in $l(\tau)$.
 The solutions are expressed in terms of
elementary exponentials (for simplicity, we write the solution in the gauge
$l(\tau) \equiv 1$ but all the results are actually gauge invariant):
\be
e^{-q_n}={|\tilde{g}_n|\over 2\mu^2_n} \biggl[ e^{\mu_n (\tau-\tau_n)}
+ e^{-\mu_n(\tau-\tau_n)} +2\ve_n \biggr] ,
\label{eq:21}
\ee
where $\tilde{g}_n = \ve \lambda_n g_n \equiv \ve g_n/\gamma_n$,
$\ve_n \equiv -\tilde{g}_n /|\tilde{g}_n|$.
            The real parameters $\mu_n^2$ and $\tau_n$ are
            the integration constants.
            In what follows we analyze the solutions with $\mu_n^2 >0$.
            In this case it is sufficient to take $\mu_n >0$, as
            $q_n(\mu_n) = q_n(-\mu_n)$.
The constraint is simply $\sum \gamma_n \mu_n^2 = 0$,
and its solution is trivial. The space of the solutions is thus defined
by the $(2N-2)$-dimensional moduli space
(one of the $\tau_n$ may be fixed)\footnote{
Of course, (\ref{eq:21}) is a solution of the 1+1 dimensional
equations that may be obtained from eqs. (\ref{eq:13}),
(\ref{eq:15b})    by choosing constant moduli
$\mu_n = \nu_n$ satisfying the
constraint (\ref{eq:15a}). Then also the moduli
$\hat{\xi}_k$ and $\hat{\eta}_k$ are constant and equal.
If the vectors $\hat{\xi}$ and $\hat{\eta}$ are constant but not
equal, the corresponding solution of the 1+1 dimensional theory given by
(\ref{eq:15c}) - (\ref{eq:15l}) may be interpreted (for $\epsilon_n > 0$)
in terms of ingoing and outgoing localized (`soliton' - like) waves
of scalar fields. We will discuss these interesting new solutions in
a separate publication.}.
            The solutions with $\mu_n^2 >0$ may have horizons.
            In fact, they have at most two horizons, and the space
of the solutions with horizons has dimension $N-1$. There exist
integrable models having solutions with horizons and no singularities but
their relation to the high dimensional world is at the moment not clear.

To prove these statements one should analyze the behavior of $q_n$ for
$|\tau| \lim \infty$
and for $|\tau - \tau_n | \lim 0$ (if $\ve_n < 0$). The horizons appear when
$F \lim -\infty$ while $\f$ and $\psi_n$ for $n \geq 2$ tend to finite
limits. This is possible for $|\tau | \lim \infty$ if and only if
            $\mu_n = \mu$.
When $F \rightarrow F_0$ and $\f \rightarrow \infty$
while the scalar matter fields are finite we have
the flat space limit, e.g. the asymptote of the BH.
            This is possible if and only if $\mu_n = a_n \mu$.
The singularities in general appear
for $|\tau - \tau_n | \lim 0$ if $\ve_n < 0$.
The scalar fields and the dilaton may be free
of singularity if and only if all $\tau_n$ are equal.

These statements may be verified if we first
write the expression for $q_n(\tau)$ in the form
\be
-q_n(\tau) = \mu_n|\tau - \tau_n| + \ln(|\tilde{g}_n|/2\mu_n^2) +
\ln[1 + \ve_n \exp (-\mu_n|\tau - \tau_n|)]^2
\label{eq:21a}
\ee
that directly follows from (\ref{eq:21})
            for $\mu_n >0$, and then look
at the corresponding analytic expressions for $F$,
$\f$ and $\psi_n$ (for $n \geq 3$) that can be derived by using
(\ref{eq:12aa}). For $|\tau| \rightarrow \infty$ we have the
following asymptotic behaviour
\be
\psi_n = -\epsilon_n \sum a_{nm} \gamma_m \{ \mu_m |\tau - \tau_m| +
\ln (2 \tilde{g}_m/\mu_m^2) \} + o(1) \ .
\label{eq:21b}
\ee
In particular,
\be
F = 2 \sum a_n \gamma_n \{ \mu_n |\tau - \tau_n| +
\ln(2 \tilde{g}_n/\mu_n^2) \} + o(1)  ,
\label{eq:21ba}
\ee
\be
\f = 2 \sum \gamma_n \{ \mu_n |\tau - \tau_n| +
\ln(2 \tilde{g}_n/\mu_n^2) \} + o(1)
\label{eq:21bb}
\ee
The divergent parts of these functions for $\tau \rightarrow \infty$
are simply
\be
\psi_n = -\epsilon_n |\tau| \sum a_{nm} \gamma_m \mu_m , \,\,
F = 2|\tau| \sum \gamma_n a_n \mu_n , \,\,
\f =  2|\tau|\sum \gamma_n \mu_n \ .
\label{eq:21bd}
\ee

When $\ve_n < 0$ and all $\tau_n$ are equal
(we may choose $\tau_n = 0$) the divergent parts of these functions are
            ($\epsilon_1 = -1, \epsilon_n = 1, n \geq 2$)
\be
\psi_n = -\epsilon_n \ln\tau^2 \sum a_{nm} \gamma_m \ , \,\,\,\,\,
F = 2 \ln\tau^2 \sum \gamma_n a_n \ , \,\,\,\,\,
\f = 2 \ln \tau^2 \sum \gamma_n \ .
\label{eq:21c}
\ee
Using the sum rules (\ref{eq:12c}) one can now
prove the above statements.
            In particular, we see that $\psi_{n \geq 3}$ and $\f$
            are finite for $|\tau| \rightarrow \infty$ if $\mu_m \equiv \mu$
            and finite for $|\tau| \rightarrow 0$ if $\tau_n \equiv \tau_0$.
            On the other hand, $F$ is singular in both these limits, its
            divergent parts being $F = -\mu |\tau|$ and $F = -\ln \tau^2$,
            respectively.
Obviously the horizons
appear when $|\tau| \rightarrow \infty$.
Thus we have the structure of the horizons and of the singularity
similar to that of the Reissner - Nordstr$\o$m BH\footnote{
Let us emphasize that when we have scalar fields varying
in space the horizons do not appear in the standard
Einstein - Maxwell theory. Thus the derived black holes are not
identical to the standard ones.
These new BH were earlier obtained in string theory and supergravity
by using other approaches  (see e.g. reviews \cite{St} - \cite{Iv}).}.
To get the black hole of the Schwarzschild type, i.e.
with one horizon, one has to
use a limiting procedure that will be discussed elsewhere.

We have seen that integrability of the theory and
asymptotic behaviour of the solutions depend only
on the $a_{mn}$. Geometric and physical features of
the solutions may depend also on the coupling constants
$g_n$. We illustrate this dependence by deriving
the two dimensional curvature. To simplify our formulas
we use the light cone coordinates $(u, v)$ and the `Weyl frame',
in which there is no
$(\nabla \f)^2 $ term in ${\cal L}^{(2)}$ and
thus no $\dot{\f}^2$ term in ${\cal L}^{(1)}$.
Then the two dimensional Ricci curvature for the
1-dimensional solutions may be derived by using
a very simple formula:
\be
R = {1\over f} \p_u \p_v \ln |f| \, = \, {1\over h(\tau)} \p_{\tau}^2 \ln |h(\tau)| \,
\equiv \,\,  \ve e^{-F} \ddot{F}(\tau) .
\label{eq:21d}
\ee
Using this formula and the expressions for $F$ derived
above one may easily prove the following statements.
1.~If $\mu_n \equiv \mu$ the curvature $R$ is finite on the
horizons, i.e. for $|\tau| \rightarrow \infty$. If in addition
all moduli $\tau_n$ are equal, $R$ is given by the following
very compact expression:
\be
R \rightarrow  -\ve_0 \prod |\tilde{g}_n|^{-2 \gamma_n a_n} \ , \,\,\,\,\,
|\tau| \rightarrow \infty ,
\label{eq:21e}
\ee
where it is supposed that all $\ve_n$ are equal and
$\ve_n \equiv \ve_0$ .
2.~If all $\ve_n$ in eq.(\ref{eq:21}) are negative
($\ve_n \equiv \ve_0 = -1$) and
$\tau_n \equiv \tau_0$, then  the curvature $R$ approaches the
finite value given in eq.(\ref{eq:21e}) also for
$|\tau - \tau_0| \rightarrow 0$. In this case $\mu_n$ may be
arbitrary real constants.
3.~If $\mu_n = a_n \mu$ we find that $R \rightarrow 0$
for $\tau \rightarrow \infty$.

Note that the solution (\ref{eq:21}) is written in a rather unusual
coordinate system. One may write a more standard representation remembering
that the dilaton $\f$ is related to the coordinate $r$ (see (\ref{eq:4})).
This may be useful for a geometric analysis of some simple solutions
(e.g. the Schwarzschild or the Reissner - Nordstr$\o$m black holes)
 but in general
the standard representation is rather inconvenient in analyzing the solutions
of the $N$-Liouville theory.

\section{Discussion and outlook}
The explicitly analytically integrable models presented here
may be of interest for different applications.
Most obviously we may use them to construct first approximations to generally
non integrable theories. Realistic theories describing BHs and CMs
            are usually not integrable.
However, explicit general solutions of the integrable approximations
may allow one to construct different sorts of perturbation theories.

For example, spherically symmetric static BHs non minimally coupled to
scalar fields are described by the integrable 0+1 dimensional $N$-Liouville
model. However, the corresponding 1+1 theory is not integrable
because the scalar coupling potentials $Z_n$ are not constant
(see eq.(\ref{eq:5})). To obtain approximate analytic solutions of the
1+1 theory one may try to approximate $Z_n$ by properly chosen constants.

This approach may be combined with the recently proposed analytic perturbation
theory allowing to find solutions close to horizons
for the most general non integrable 0+1 DG theories \cite{atfm}.
Near the horizons we can also use the integrable 1+1 dimensional
DG (with $Z_n = -1$) as a good approximation to a realistic
theory (with $Z_n$ depending on $\f$).

In cosmological applications, the behaviour of the
1+1 dimensional solutions for $\f \rightarrow 0$ (i.e. near
the singularity at $\f =0$) is of great interest.
Integrable 1+1 dimensional theories could give, at best,
a rough qualitative approximation of the exact solutions near
the singularity. A more quantitative approximation might be
obtained by first asymptotically solving the exact theory
in the vicinity of $\f =0$ and then sewing the asymptotic
solutions with those of the  integrable theory. To realize
such a program one needs a very simple and explicit analytic
solutions of the integrable theory. Our simple model having
the solutions represented in terms of the moduli $\hat{\xi}$
and $\hat{\eta}$ may give a good starting point for such a work.
Of course, before applications to realistic cosmologies become
possible, one should study in detail and completely classify
and interpret the behaviour of the 1+1 dimensional solutions
and their precise relation to the 1+0 dimensional reduction.

The reduction from dimension 1+1 both to dimension 1+0 and to
dimension 0+1 is especially transparent in the moduli representation
for the solutions of the 1+1 dimensional $N$-Liouville model
However, as we emphasized above, the whole procedure of the
dimensional reduction should be reconsidered from a more general
point of view. A more detailed motivation for reconsidering the
usual procedures will be given in a forthcoming paper\cite{VDA2}.

\newpage

\section{Appendix}

\subsection{Reduction of the Curvature}

Let the block diagonal dimensional reduction be
$ ds^2 = g_{ij} dx^i dx^j + h_{mn} dx^m dx^n ,$
where the metric $g_{ij}$  depends only on the coordinates of the first
subspace, $x^i$.
The Ricci curvature scalar for this metric is then
\ba
R = R[g] + R[h] &-&{2 \over \sqrt{h}} \nabla^m \nabla_m \sqrt{h} +
\quart g^{ij} \p_i h^{mn} \p_j h_{mn} + \nonumber \\
&+& \quart g^{ij} (h^{mn} \p_i h_{mn})\, (h^{pq} \p_j h_{pq})  \, .
\label{a2}
\ea
Using this expression, partial integrations, and the Weyl transformations
one may easily derive the reductions presented in the main text.
If the second subspace is a $(d-2)$-sphere $S^{(d-2)}(k)$
of radius $e^{\beta}$ then
   $$R[h] = R[S^{(d-2)}(k)] = e^{-2\beta} k(d-2)(d-3) \, , \,\,\,\,
    k=\pm 1,\, 0 \, .$$
For the analysis of the  geometric properties of different 4-dimensional
cosmologies  it may help to use the easily derivable formula for the
scalar curvature of the 3-dimensional spherically symmetric subspace
of the space with the metric (\ref{eq:4a}) (with $d=4$, $k=1$ and fixed $t$)
\be
R = R^{(3)} = 2e^{-2\beta} -2 e^{-2\alpha}(2{\beta}^{\prime \prime} +
3{{\beta}^{\prime}}^2 - 2{\beta}^{\prime}{\alpha}^{\prime}) \,\,\,
\label{eq:a3}
\ee
To analyze the geometric properties of the 3-space in more
detail one may use the expressions for its Ricci tensor
\be
R_1^1 = -2e^{-2\alpha}({\beta}^{\prime \prime} +
{{\beta}^{\prime}}^2 - {\beta}^{\prime}{\alpha}^{\prime}) \, , \,\,\,\,
R_2^2 = R_3^3 = \half (R^{(3)} - R_1^1) \, .
\label{eq:a4}
\ee

To help the reader in keeping  trace of relations between the dimensions $d$,
1+1, 1+0 and 0+1 we
also write here a simple expression for the curvature in dimension 1+1.
Given the metric in diagonal form as
$ds^2 = -e^{2\gamma}dt^2 + e^{2\alpha}dr^2 $,
its Ricci scalar $R$ is
\be
R = 2 e^{-2\gamma} (\ddot{\alpha} + {\dot{\alpha}}^2 - \dot{\alpha}
\dot{\gamma} ) -
2 e^{-2\alpha} (\gamma^{\prime\prime} + {\gamma^{\prime}}^2 -
\gamma^{\prime} \alpha^{\prime}) .
\label{a5}
\ee
Further, for any scalar field $\f$ the expression for $\nabla^2 \f$ is
\be
\nabla^2 \f \equiv \nabla^m \nabla_m \f = - e^{-2\gamma} \bigl( \ddot{\f}
+(\dot{\alpha} - \dot{\gamma}) \dot{\f}) + e^{-2\alpha}
({\f}^{\prime\prime} +
({\gamma}^{\prime} - {\alpha}^{\prime}) {\f}^{\prime} \bigr) .
\label{a6}
\ee

All these expressions simplify in the $(u,v)$ coordinates that can be
obtained by taking $\gamma = \alpha$ and introducing the light cone
coordinates (always possible for the 1+1 metric having the Minkowski signature).
Denoting $e^{2\gamma} = e^{2\alpha} = f$ we have $ds^2 = f(dr^2 - dt^2)$,
$ R =  f^{-1} (\p_t^2 - \p_r^2) \ln{|f|}$.
The $(u,v)$ metric, which drastically simplifies the
equations of motion and all computations, can be obtained e.g. writing
 $t = u + v$ and $r = u - v$: $ds^2=-4\, f(u,v) \, du \, dv$ .

There is a residual symmetry in the $(u,v)$ coordinates, namely,
$u \rightarrow a(u)$,
$v \rightarrow b(v)$. Under this transformation the previous equation becomes
\be
ds^2=-4\, f(a(u),b(v)) \, a'(u) \, b'(v) \, du \, dv \, =
-4\, f(a,b) \, da \, db .
\label{a9}
\ee
Thus the metric in the coordinates $(a,b)$ is the same as in the
$(u,v)$ coordinates.
Also the curvature and equations of motion remain invariant.

This freedom is useful for many reasons. For example, suppose we have found
a solution of the EMs for which the metric $f$ and the dilaton
$\varphi$ depend only on $uv$. Then, choosing $a = \ln{u}$, $b = \ln{v}$, we
may go to coordinates $(a,b)$ in which the metric function and the dilaton
depend on $a + b$.\footnote{
This means that the 1+1 metric,
            as well as HD spherical metric (\ref{eq:5a}) from which it originated,
            is effectively 1-dimensional.
This, however, does not mean that
the whole theory reduces to the dimension one, because the scalar matter fields
may still depend on two variables \cite{FI}.}
More generally, the solutions of integrable models may usually be written in terms
of massless free fields $\chi_n$ which are solutions of the D'Alembert equation
and thus may be written as $\chi_n(u,v) = a_n(u) + b_n(b)$. If all $\chi_n$
are equal, i.e. $\chi_n = a(u) + b(v)$, the theory reduces to one dimension.

In the same way one may dimensionally reduce the general, non integrable models.
We describe the simplest approach using the light cone coordinates. The more
standard approach that uses $r$ and $t$ is more cumbersome but may be of use
in interpreting the low dimensional solutions as solutions of higher
dimensional theories.
As emphasized in the main text, the described naive reduction does not
allow to obtain some physically interesting one-dimensional solutions
of higher dimensional gravity coupled to matter fields. It seems that a
more general approach should use the dimensional reduction of the equations
of motion, in other words, it should employ a general procedure for
separation of the time and space variables.

\subsection{Reduction of the Equations}

The equations of motion for consistently reduced gravity theories
 are equivalent to the Einstein equations and there is no loss of information.
 Having the complete set of the equation one may use any convenient gauge
(coordinate system).
Let us write the EMs in the light cone ($u,v$) coordinates. To simplify
the formulas we keep only one scalar field and thus use, instead of (\ref{eq:7}),
the following effective Lagrangian
\be
{\cal L}^{(2)} = \sqrt{-g}\, \biggl[ \f R + V(\f,\psi) +
                    \sum Z_n(\f ,\psi)(\nabla \psi_n)^2 \biggr].
\label{a11}
\ee
By first varying this Lagrangian in generic coordinates and then
going to the light cone ones we obtain the equations of motion
 \be
 \p_u \p_v \f+f\, V(\f,\psi)=0, \label{F.15}
 \ee
  \be
  f \p_i ({{\p_i \f} \over f }) \, = \sum Z_n (\p_i \psi_n)^2  \,\,\,\,\,\,\,\,\,
 (i=u,v)
 \label{F.17}
 \ee
 \be
\p_v (Z_n \p_u \psi_n) +\p_u (Z_n \p_v \psi_n) + f V_{\psi_n}(\f,\psi)=
\sum Z_{m, \psi_n} \, \p_u \psi_m \, \p_v \psi_m \ ,
 \label{F.16}
 \ee
 \be
 \p_u\p_v\ln |f| + f V_{\f}(\f,\psi) = \sum Z_{n, \f} \,\p_u \psi_n \, \p_v\psi_n \ ,
 \label{F.18}
 \ee
 where $V_{\f}= \p_{\f} V$, $V_{\psi_n} = \p_{\psi_n} V$,
 $Z_{n, \f}= \p_{\f} Z_n$, and $Z_{m, \psi_n} =\p_{\psi_n} Z_m$.
These equations  are not independent. Actually,
(\ref{F.18}) follows from (\ref{F.15}) $-$  (\ref{F.17}). Alternatively,
if  (\ref{F.15}), (\ref{F.17}), (\ref{F.18}) are satisfied,
one of the equations (\ref{F.16})
is also satisfied.

The most important equations are the constraints (\ref{F.17}).
A general formulation of the (naive) dimensional reduction in the $(u,v)$
coordinates that is valid both for static and cosmological solutions
is suggested by the following simple observation.
Consider the solutions with constant scalar
field $\psi \equiv \psi_0$ (the `vacuum' solution). This solution exists if
 $V_{\psi}(\varphi,\psi_0) = 0$, see eq.(\ref{F.16}).
 The constraints (\ref{F.17}) can now be solved because their right-hand
sides are identically zero. It is a simple exercise to prove that there exist chiral
fields $a(u)$ and $b(v)$ such that $\varphi (u,v) \equiv \varphi (\tau)$
and $f(u,v) \equiv \varphi^{\prime} (\tau) \, a'(u) \, b'(v)$
(the primes denote derivatives with respect to the corresponding
argument). Using this result it is easy to prove that eq.(\ref{F.15})
has the integral $ \varphi^{\prime} + N(\varphi) = M $,
where $N(\varphi)$ is defined by the equation $N^{\prime}(\f) = V(\f , \psi_0)$
and $M$ is the integral of motion which for the BH solutions is
proportional to the mass of the BH. The horizon, defined as a zero
of the metric $h(\tau) = M - N(\varphi)$, exists because the equation
$M = N(\varphi)$ has at least
one solution in some interval of values of $M$. The EMs
in the case considered are actually dimensionally reduced. Their solutions
can be interpreted as BHs (Schwarzschild, Reissner Nordstr{\o}m
and other known BH solutions in any dimension) or as CMs.

Taking into account the lesson of the scalar vacuum solutions, we introduce
a more general dimensional reduction by supposing that the scalar fields and
the dilaton depend
on one free field $\tau = a(u)+b(v)$
(after dimensional reduction it is interpreted either
as the space or the time coordinate).
Then from eq.(\ref{F.15}) it follows that the metric should have the form
$ f(u,v) =\ve \, h(a+b)\, a'(u) \, b'(v)$,
where $\ve$ is introduced in order to have the same type of metric (\ref{eq:17})
for the 0+1 and 1+0 cases. Defining
$ \tau\equiv a+b$, $ \bar \tau\equiv a-b$,
we have
\be
 ds^2=-4f(u,v) \, du \, dv\,=\,-4\ve h \, da \, db\,=\,
 -\ve h \bigl(d\tau^2-d \bar \tau^2\bigr) \, , \label{a15}
\ee
and thus both reduced metrics may be written as (\ref{eq:17}) by choosing
 $\tau=r$, $\ve=-1$ or $\tau=t$ and $\ve=+1$.

 The reduced EMs for the dilaton and the scalar field\footnote{
 To simplify notation and formulas we
 consider here only the case of one scalar
 field. Generalizing to any number of fields is obvious.},
\bdm
\p_{\tau}^2 \f +\ve hV=0 \, ,  \qquad
2 \p_{\tau} \bigl( Z\p_{\tau}\psi\bigr) +\ve h V_{\psi} = Z_{\psi}
\bigl(\p_{\tau}\psi\bigr)^2 \, ,
\edm
depend on $\varepsilon$ while the constraints are the same for both reductions
and give just one reduced constraint,
$ \p_{\tau}^2 \f - \p_{\tau}\f \, \p_{\tau} \ln|h| =
Z(\p_{\tau}\psi)^2 \,$ ,
equivalent (in the standard terminology) to the energy constraint.

Thus we have {\it the rule} for the reduction of the EMs:
using the equations in the light cone gauge, derive the equations for
$\f(\tau)$, $\psi(\tau)$, $\ve h(\tau)$ and then take
$\tau=r$ and $\ve=-1$  or $\tau=t$  and $\ve=+1$.
Using this rule we may write down reduced equations without calculations.
First take the gauge fixed Lagrangian
in the $(u, \ v)$ metric and using
the residual covariance with respect to the
transformation $u\ra a(u)$, $v\ra b(v)$
transform it to the new coordinates $(a, \ b)$,

\bdm
{\cal L}_{\rm g.f.}^{(2)} =\f \, \p_u\p_v F +f V-Z \, \p_u \psi \, \p_v\psi \mapsto
\f \p_a \p_b F +\ve\, h V - Z \, \p_a \psi \, \p_b \psi \
\edm
(recall that  $F=\ln |f|$).
Then, using the reduction rule, we obtain
$${\cal L}_{\rm g.f.}^{(1)} = \f \ddot F  -Z\dot \psi ^2 + \ve h V ,$$
where the dot denotes $\tau$-differentiation.
To get rid of the second derivative,
we neglect the total derivative in the Lagrangian
and replace $\f \ddot F$ by $\dot \f \dot F$.
This Lagrangian gives the correct gauge fixed equations of motion.
To restore the lost constraint (the gauge fixed Lagrangian
does not give the constraints) we recall that the constraint
is just ${\cal H}=0$, where ${\cal H}$ is the Hamiltonian correspondent
to the Lagrangian. It is evident that
${\cal H} =-\dot \f \dot F -Z\dot \psi -\ve h V$.
Now it is easy to guess that the correct Lagrangian giving the equations
of motion and the constraint ${{\cal H} = 0}$ is simply
\be
{\cal L}^{(1)} = - {1\over l(\tau) } \biggl[ \dot\f \dot F + Z\dot \psi^2 \biggr]
+ l(\tau) \ve h V.
\label{a24}
\ee
In order to obtain from here the 0+1 theory we simply take $\tau=r$ and
$\ve=-1$. The 1+0 theory can be written taking $\tau=t$ and $\ve=+1$.

Finally, let us write an example of cosmological reduction directly from
a higher dimensional theory. We take the $d$-dimensional metric (\ref{eq:5a}),
suppose that the scalar functions depend on one variable $t$,
see e.g. \cite{St}. Then using eq.(\ref{a2}) with
the 1-dimensional metric
            $g_{ij}$ and the $(d-1)$-dimensional metric $h_{mn}$ we can find
for example the reduced action for the Lagrangian (\ref{eq:3b}).
We write here only the reduced curvature part (the derivation of the other terms
is obvious):
\ba
S&=& \int d^d x \, \sqrt{-g} \, \sqrt{h} \, R^{(d)}\nonumber \\
&=& (d-1)(d-2) \int dt \, e^{\gamma} \, e^{\alpha (d-1)} \bigl[ \, k e^{-2\alpha}
- e^{-2\gamma} {\dot{\alpha}}^2  \bigr] .
\label{a25}
\ea

It is not difficult to check that the naive cosmological reduction of
the 1+1 theory does not give this expression for the cosmological action
(see \cite{VDA2}). The geometric explanation of this fact is given by
 eq.(\ref{eq:a4}), from which it is clear that for ${\beta}^{\prime}=0$
we have $R_1^1 \neq R_2^2$ and thus the 3-space is not isotropic while
for all three cosmologies described by this action it must be isotropic.
To obtain these standard cosmologies one has to use a different
dimensional reduction from 1+1 to 1+0 dimension in which both the
metric and the dilaton may depend on $r$ and $t$.
            Indeed, according to (\ref{eq:a4}), the isotropy condition
            $R_1^1 = R_2^2$ requires that
\be
e^{2(\beta - \alpha)} ({\beta}^{\prime \prime} - \beta^{\prime} \alpha^{\prime}) = -1
\label{eq:a26}
\ee
            This may be satisfied
if we suppose that $\alpha = \alpha_0(t) + \alpha_1(r)$
 and $\beta = \beta_0(t) + \beta_1(r)$. Then it follows that
 $\alpha_0(t) - \beta_0(t) = {\rm const}$ and, choosing the gauge in which
 $\alpha_1(r) = {\rm const}$, one can show that the solutions, up to a
 gauge choice, may be written as $e^{\beta_1} = \sinh(\sqrt{k} r)/\sqrt{k}$,
 where $k = 0, \pm 1$. This gives the standard cosmological solutions from
 the dimensional reduction of the 1+1 dimensional dilaton gravity. The detailed
  derivation of the reduced Lagrangian and HD equations of motion will be
presented in \cite{VDA2}.

\subsection{Nonlinear coupling of gauge fields}
Suppose that in place of the standard Abelian gauge field term,
$X(\f,\psi) F^2$, the Lagrangian contains a more general coupling of
        $A$ gauge fields $F_{ij}^{(a)}=\p_i A_j^{(a)} -\p_j A_i^{(a)}$
        to dilaton and scalar fields, $X(\f, \psi; F^2_{(a)})$,
        where $F^2_{(a)} \equiv g^{ik} g^{jl} F_{ij}^{(a)} F_{kl}^{(a)}$
        and $a = 1,...,A$\footnote{
        In what follows we denote the set of all $F^2_{(a)}$ in the potential
        $X$ simply by $F^2$, i.e. $F^2 = \{F^2_{(1)},...,F^2_{(A)}\}$ and thus
        $X = X(\f, \psi; F^2)$.}.
        Without loss of generality we may write the
        two dimensional Lagrangian as
\be
{\cal L}^{(2)} = \sqrt{-g}\, \biggl[ \f R + X(\f, \psi; F^2) +
    \sum Z_n(\f ,\psi)(\nabla \psi_n)^2 \biggr].
\label{e11}
\ee
        As in the case of the Lagrangian (\ref{a11}) the equations of motion
        are obtained by varying the Lagrangian (in generic coordinates)
        with respect to all variables, including now the gauge fields, and then going
        to the light cone coordinates (\ref{eq:16}). Variations of $\psi_n$ and
        of $\f$ give the equations (\ref{F.16}) and (\ref{F.18}) with $V$ replaced
        by $X$. The equations (\ref{F.15}) and (\ref{F.17}) are obtained by varying
        the metric. The two equations (\ref{F.17}) correspond to variations of the diagonal
        part $g^{ii}$. It is not difficult to see that the additional term produced
        by the $F^2$ dependence of $X$ vanishes when we pass to the light cone metric.
        Indeed, this contribution is proportional to the sum of the terms
        $(\p X / \p F^2_{(a)}){\delta F^2_{(a)} / \delta g^{ii}}$ and this vanishes in the light
        cone metric because
        ${\delta F^2 / \delta g^{ii}} = 2 g^{jk} F_{ij} F_{ik}\equiv 0$ if
        $g_{jj} = 0$. In other words, the constraint equations (\ref{F.17}) are
        insensitive to the $F^2$ dependence of $X$ in the $(u,v)$ coordinates.
        In contrast, the equation (\ref{F.15}) has additional terms even in the
        light cone coordinates:
\be
 \p_u \p_v \f + f \biggl[ X(\f,\psi ; F^2) +
 f \sum {\p X \over \p F^2_{(a)}} \cdot {\p F^2_{(a)} \over \p f} \biggr] = 0.
 \label{e15}
 \ee

        Finally, we have the equations of motion for the gauge fields,
\be     \label{F.01}
\p_i \biggl( \sqrt{-g} \, {\p X \over \p(\p_i A_j^{(a)})} \biggr) =0 ,
\ee
where ${\p X / \p (\p_i A_j^{(a)})} = 4 F^{ij}_{(a)} {\p X / \p F^2_{(a)}}$.
        In dimension 1+1 the equations (\ref{F.01}) reduce to the conservation laws
 \be
 \sqrt{-g} \, F^{ij}_{(a)} {\p X \over \p F^2_{(a)}} = \varepsilon ^{ij}
 \lambda Q_a
 \label{F.03}
 \ee
 where $\ve^{ij}=-\ve^{ji}$, $\ve^{01}=1$, $\lambda$ is a constant to be
 defined later and $Q_a$ are conserved charges.
 From this it is easy to obtain the equations for $F^2_{(a)}$ (recall that
$2g=\ve^{ij} \ve^{lk} g_{il} g_{jk}$):
\be  \label{F.04}
F^2_{(a)}  = -2 \lambda^2 \,Q^2_a \biggl( {\partial X
\over \partial F^2_{(a)}} \biggr)^{-2}.
\ee
This allows us (in principle) to write $F^2_{(a)}$ (and  $F^{ij}_{(a)}$) in terms of
$\f,\,\psi,\, Q$,
        where $Q$ is the set of all the charges, $Q = \{Q_1,...,Q_A\}$.
        Let us denote the solution as
        $\bar F^2_{(a)} \equiv \bar F^2_{(a)}(\f,\psi;Q)$ (or simply $ \bar F^2_{(a)})$.
Now we can write $\bar F^{ij}_{(a)} $ in terms of $\bar F^2_{(a)}(\f,\psi;Q)$.
Eq. (\ref{F.04}) gives
\be
{\p \bar X \over \p \bar F^2_{(a)}} \equiv {\p X \over \p F^2_{(a)}}({F^2}
\Rightarrow {\bar F^2})
 = \epsilon_a \sqrt{2}{ \lambda Q_a \over {\sqrt{-\bar F^2_{(a)}}}} ,
\label{F.05}
\ee
where $\bar X \equiv X(\f, \psi; \bar F^2)$,
$\epsilon_a = {\rm sign} [ \, {\p X / \p F^2_{(a)}} ]$ and
$\sqrt{-F^2_{(a)}} > 0$\footnote{
For small values of $F^2_{(a)}$ we usually have $\epsilon_a < 0$. }.
Then from the equations (\ref{F.03}) and (\ref{F.05}) we get
 \be  \label{F.06}
 \bar F^{ij}_{(a)} = {\ve^{ij} \lambda Q_a \over \sqrt{-g}} \, \biggl( {\p \bar X
 \over \p F^2_{(a)}} \biggr)^{-1} = \ve^{ij}\,\epsilon\, \, {\sqrt{ \bar F^2_{(a)} / 2g }} .
 \ee

        Now we can exclude the gauge fields from the equations of motion
        and find an effective potential $X_{\rm eff}$
        depending only on $\f,\psi$ and $Q$.
        To do this we go to the $(u,v)$ coordinates and recall that
        {\it after} computing the constraints, which are insensitive to
        the $F^2$ dependence, one may derive other equations using
the reduced (gauge fixed) Lagrangian
\be   \label{F.08}
{\cal L}_{\rm g.f.}^{(2)} = \f \p_u \p_v  \ln |f|  + fX(\f,\psi;F^2) -
 \sum Z_n \, \p_u \psi_n \, \p_v \psi_n \, .
\ee

        It is not difficult to guess that if we exclude the dependence on
        $F^2$ from the expression in the square brackets in eq.(\ref{e15})
        we will get the desired effective potential $X_{\rm eff}$ depending
        on $\f$, $\psi$ and $Q$. To write the explicit expression for it
        we first note that (we don't set yet $F^2=\bar F^2$)
\be  \label{F.09}
 X \,+\, f \sum {\p X\over \p F^2_{(a)}} \cdot {\p F^2_{(a)} \over \p f} \cdot =
 X \,-\, \sum 2F^2_{(a)} {\p X\over \p F^2_{(a)}} \,
 \ee
        Replacing $F^2$ by $\bar F^2$ and using in (\ref{F.09}) the relations
        (\ref{F.05}) and (\ref{F.06}) we obtain the expression for
        the effective potential (we now choose $2\sqrt{2}\lambda = 1$)
\be  \label{F.11}
X_{\rm eff}(\f,\psi, \bar F^2) \equiv
X_{\rm eff}(\f,\psi, Q) \equiv  X(\f,\psi; \bar F^2) \,
+ \sum Q_a \epsilon_a \sqrt{-\bar F^2_{(a)}}
\ee
that should replace $X$ in the Lagrangian (\ref{F.08}).
        If we consider $X_{\rm eff}$ as a function of $\bar F^2$ we may reproduce
        the expressions for $\bar F^2_{(a)}(\f,\psi,Q)$.
        Indeed, varying the effective Lagrangian
\be   \label{e08}
{\cal L}_{\rm eff}^{(2)} = \f \p_u \p_v  \ln |f|  + fX_{\rm eff}(\f,\psi; \bar F^2)
 - \sum Z_n \, \p_u \psi_n \, \p_v \psi_n .
\ee
with respect to $\bar F^2$ we find the condition
\be
\delta {\cal L}_{\rm eff}^{(2)} / \delta \bar F^2_{(a)} =
f \biggl[ {\p X \over \p \bar F^2_{(a)}}  -
\half \epsilon Q_{(a)} / \sqrt{- \bar F^2_{(a)}} \biggr] = 0
\label{e09}
\ee
that gives for $\bar F^2_{(a)}$ the expression defined by (\ref{F.04})
        (with $\lambda^2 = 1/8$).

 If these conditions are satisfied, the total derivative of $X_{\rm eff}$ in $\f$
 coincides with the partial derivative of $X$ (with fixed $F^2$).
 Indeed, we have
\bdm
{d X_{\rm eff} \over d\f}={\p X\over \p \f}  +
\sum \biggl[ {\p X \over \p \bar F^2_{(a)}}  -
 {\half \epsilon  Q_{(a)} / \sqrt{- \bar F^2_{(a)}}} \biggr]
{\p \bar F^2_{(a)} \over \p \f }
=  {\p X\over \p \f}
 \edm
 due to (\ref{F.05}). Analogously,
${d X_{\rm eff} / d\psi}={\p X / \p \psi}$.
This means that the equations (\ref{F.16}), (\ref{F.18})
obtained from the effective Lagrangian (\ref{e08})
 by varying $\psi, \f$
coincide with the corresponding equations obtained from the original
Lagrangian ${\cal L}^{(2)}$.
The constraints (\ref{F.17}) are obviously the same and variation
of $\f$ gives the correct equation (\ref{e15}).
It follows that we may forget about the fields $F_{ij}$, $A_i$
(that can be derived from eq.(\ref{F.06}) if needed)
and work with the effective theory for the fields $f, \f, \psi$
simply replacing the original Lagrangian (\ref{F.08}) with
the effective Lagrangian.

 It is not difficult to apply this construction to
known Lagrangians of the Dirac - Born - Infeld type as well as
to find new integrable models with nonlinear coupling of Abelian gauge
fields to gravity.

\bigskip
\bigskip

 {\bf Acknowledgment:} For financial support and kind hospitality one of the authors (ATF)
 is grateful to the Dept. of Theoretical Physics of the University of Turin and
 INFN (Section of Turin), to CERN-TH, to the MPI Muenchen and W.Heisenberg
 Institute, where some results were obtained. Useful discussions with
 P.~Fr\'e, D.~Luest, D.~Maison and G.~Veneziano are kindly acknowledged.
                This work was also partly supported by RFBR grant 03-01-00781-a.

\end{document}